# Multi-Resolution 3D CNN for MRI Brain Tumor Segmentation and Survival Prediction

Mehdi Amian[1] and Mohammadreza Soltaninejad[2]

[1]Control and Intelligent Center of Excellence, School of Electrical and Computer Engineering,
University of Tehran, Iran
[2]School of Computer Science, University of Nottingham, UK
`mehdi.amian@ut.ac.ir, m.soltaninejad@nottingham.ac.uk`

**Abstract.** In this study, an automated three dimensional (3D) deep segmentation approach for detecting gliomas in 3D pre-operative MRI scans is proposed. Then, a classification algorithm based on random forests, for survival prediction is presented. The objective is to segment the glioma area and produce segmentation labels for its different sub-regions, i.e. necrotic and the non-enhancing tumor core, the peritumoral edema, and enhancing tumor. The proposed deep architecture for the segmentation task encompasses two parallel streamlines with two different resolutions. One deep convolutional neural network is to learn local features of the input data while the other one is set to have a global observation on whole image. Deemed to be complementary, the outputs of each stream are then merged to provide an ensemble complete learning of the input image. The proposed network takes the whole image as input instead of patch-based approaches in order to consider the semantic features throughout the whole volume. The algorithm is trained on BraTS 2019 which included 335 training cases, and validated on 127 unseen cases from the validation dataset using a blind testing approach. The proposed method was also evaluated on the BraTS 2019 challenge test dataset of 166 cases. The results show that the proposed methods provide promising segmentations as well as survival prediction. The mean Dice overlap measures of automatic brain tumor segmentation for validation set were 0.84, 0.74 and 0.71 for the whole tumor, core and enhancing tumor, respectively. The corresponding results for the challenge test dataset were 0.82, 0.72, and 0.70, respectively. The overall accuracy of the proposed model for the survival prediction task is %52 for the validation and %49 for the test dataset.

**Keywords:** Convolutional Neural Network, U-Net, Deep Learning, MRI, Brain Tumor Segmentation.

## 1 Introduction

Brain tumors are caused by abnormal growth of the cells inside brain and have a wide variety of tumor types. They can be generally categorized into low-grade gliomas (LGG) or high-grade gliomas (HGG). Magnetic resonance imaging (MRI) plays an important role regarding the clinical tasks related to brain tumors. Accurate segmen-tation of brain tumor may aid the measurement of tumor features to help diagnosis, treatment planning and survival prediction [1]. MR images can be generated using different acquisition protocols such as fluid attenuated inversion recovery (FLAIR), T1-weighted (with and without contrast agent), and T2-weighted to distinguish between different tumor sub-tissues.

Segmentation of human brain tumor in medical images is a vital and crucial task that traditionally is performed manually by physicians. The manual delineation practices are subjective and inherently prone to misinterpretation that can bring about sever and even fatal upcomings. So, developing a reliable and fast automated algorithm, undoubtedly, leads to much more accurate diagnosis, making a remarkable advance in long term in treatment planning for the patients. This becomes even more highlighted when having three-dimensional observation to images by the machine instead of a natural two-dimensional view of a human interpreter.

So far, many efforts addressed inventing such an automatic segmentation system. Undeniably, several big steps have been taken, yet there is a lot to be taken. On the other hand, thanks to emerging powerful computing processors as well as availability of big datasets, deep learning and particularly its recent advancements revolutionized many aspects of the technology by manifesting unprecedented amount of knowledge and learning about various data types, i.e. text, speech, and image.

Deep learning is widely being used in medical imaging domain in various ways such as denoising, finding biomarkers, pattern prediction, and detecting lesions and tumors. Applying deep learning techniques to multimodal MR images for tumor segmentation is naturally a challenging task due to high dimensionality of the input data, poor quality and problems of the image during capturing such as bias field, and after all designing an appropriate architecture for the specific objective.

Due to the recent advances in deep neural networks (DNN) in recognition of the patterns in the images, most of the recent tumor segmentations have focused on deep learning methods [2]. Fully convolutional networks (FCN) have been suggested for per-pixel classification with the advantage of end-to-end learning [3]. Despite the advantage of dense pixel classification, FCN-based methods suffer from the loss of spatial information, which occurs in the pooling layers, results in coarse segmentation [2]. U-Net [4] proposed using skip layers to tackle this problem and was suggested for fine medical image segmentation. Several methods have proposed using U-Net for brain tumor segmentation [5-7]. U-Net showed promising results in medical image segmentation tasks. So, several researches focused on U-Net modification to acquire even better outcomes. Cascaded U-Net is a successful example of such modifications as presented in [8] where the cascaded U-Net outperforms the standard version.

The present study on segmentation is inspired by two deep convolutional neural networks, i.e. U-Net and the one that is proposed in [9]. These networks are placed in parallel stream lines fed by original images. The U-Net is meant to capture the local features and make a fine learning of the data. On the other hand, the other pipeline is to maintain a coarse but global learning.

Furthermore, the proposed architecture is designed to take the whole image as the input for the network, rather than patch-based architectures which incorporate partial information from the images during training. This approach ensures that all the se-mantic features throughout the whole volume will be considered during training, which eliminates most of the false positives.

For the survival prediction task, a model based on the random forest (RF) [10] is presented. The output of the model is as the number of days. As an ensemble learning algorithm, RF is widely being used in both classification and regression purposes.

## 2      Materials and Methods

### 2.1    Dataset

The proposed network is trained using the Multimodal Brain Tumor Segmentation Challenge (BraTS) 2019 [11-15] training dataset which includes 259 HGG and 76 LGG patient cases. The dataset contains segmentation ground truth manually annotated by experts and provided on the Center for Biomedical Image Computing and Analytics (CBICA) portal. The network was evaluated using BRTAS 2019 validation dataset which includes 125 patient cases. For the task of post-operative survival prediction, 101 of the training patient cases were provided with the survival information. The network was also evaluated for the challenge test dataset which includes 166 patient cases for the segmentation task and 107 cases for the survival prediction task.

### 2.2    Segmentation

Our proposed segmentation method consists of two main pipelines steps with different resolution levels, i.e. original and low resolutions. The architecture of the proposed network is depicted in Fig. 1. The pre-processing stage consists of intensity normalization, histogram matching and bias filed correction. The intensities were normalized for each protocol by subtracting the average of intensities of the image and divided by their standard deviation. The histogram of each image was normalized and matched to a selected reference image, which is one of the patient cases. Bias field correction was performed using the toolbox provided in [16].

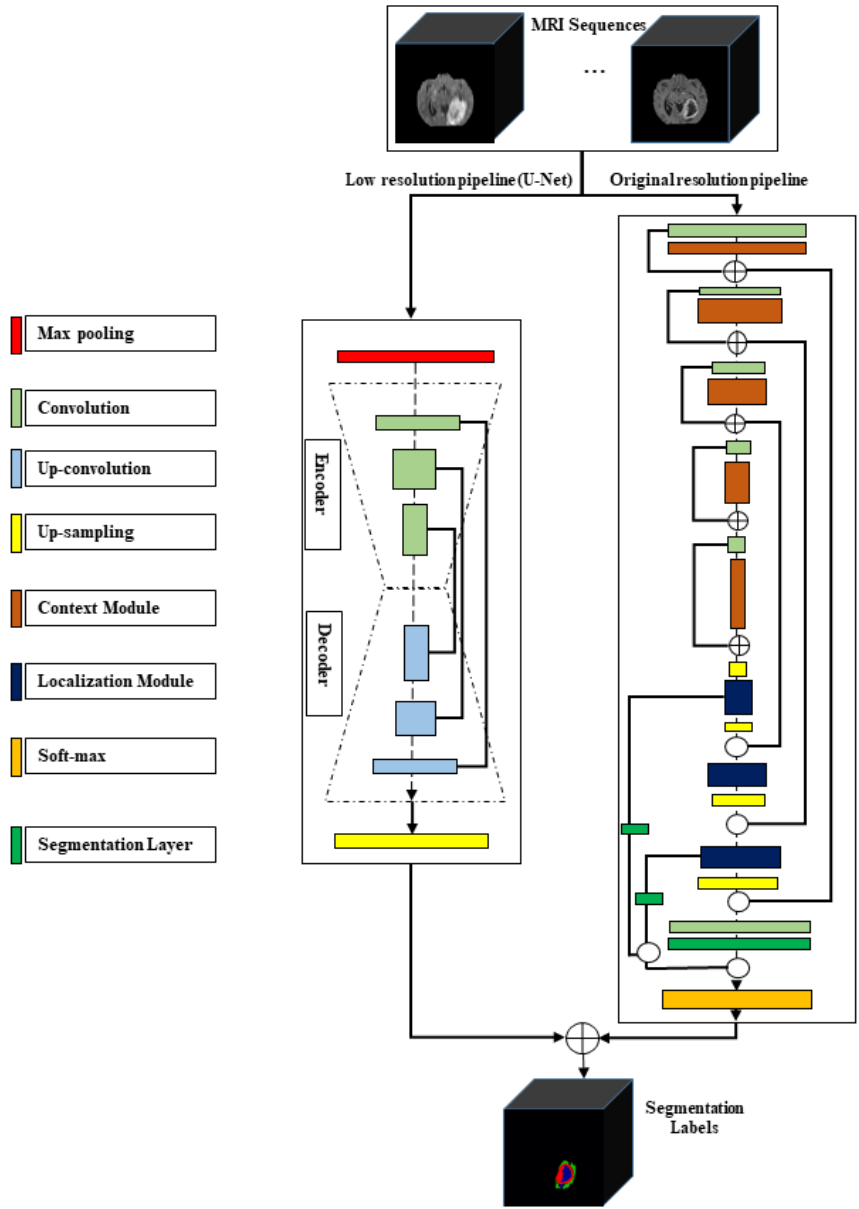

**Fig. 1.** Architecture of the proposed multi-resolution method consisting of two different pathways one with original resolution and the other with lower resolution and larger field of view.

### 2.3 Survival Prediction

The output masks forming the segmentation in the previous section are used for the task of survival prediction. Some spatial features are extracted for the whole tumor and each sub-tissue. Normalized volume size for the whole tumor, tumor core and enhanced tumor are calculated. The average intensity value for each sub-tissue is another feature. The feature vectors extracted for each volume of interest (VOI) were input to the random forests (RF). RF parameters, i.e. tree depth and the number of trees, were tuned by examining them on training datasets and evaluating the classification accuracy using 5-fold cross validation. The number of 30 trees with depth 10 provided an optimum generalization and accuracy. The RF classifier was used in regression mode which produced predictions as number of days. The schematic diagram of the proposed approach for the survival prediction is demonstrated in Fig 2.

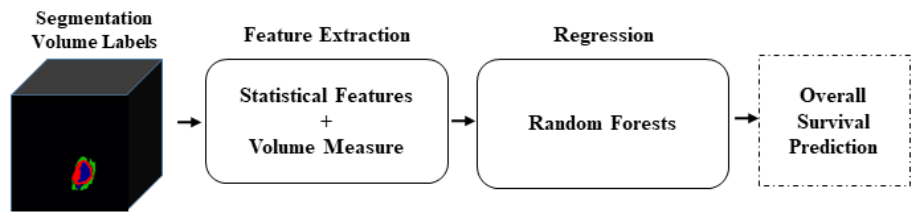

**Fig. 2.** The pipeline for overall survival prediction based on the volume labels extracted in the segmentation stage

## 3   Results

### 3.1   Segmentation Task

For the segmentation task, the proposed method was implemented using Keras Tensorflow backend on Nvidia GeForce GTX 1080 Ti GPU, RAM 11 GB, PC with CPU Intel Core i7 and RAM 16 GB with the operating system Linux. The U-Net [4] was modified and the method in [9] was implemented using [17]. The RF was implemented using MATLAB 2019a. The ground truth is provided for the training set, whilst for evaluation a blind testing system is utilized. The evaluation measures which are provided by the CBICA's Image Processing Portal, i.e. Dice score, sensitivity, specificity, Hausdorff distance, were used to compare the segmentation results with the gold standard (blind testing).

Table 1 presents the evaluation results obtained by applying the proposed segmentation method on BraTS 2019 validation dataset which was provided by CBICA blind testing system. Figure 3 shows segmentation results of the proposed multi-resolution approach for some cases of BraTS 2019 training dataset and the ground truth. Two modalities, i.e. FLAIR and T1-ce, are shown in Fig 2 and the tumor sub-tissues are overlaid on T2 modality and depicted in axial, sagittal, and coronal views. Figure 4 represents segmentation results of the proposed multi-resolution method for two sample cases of BraTS 2019 validation dataset.

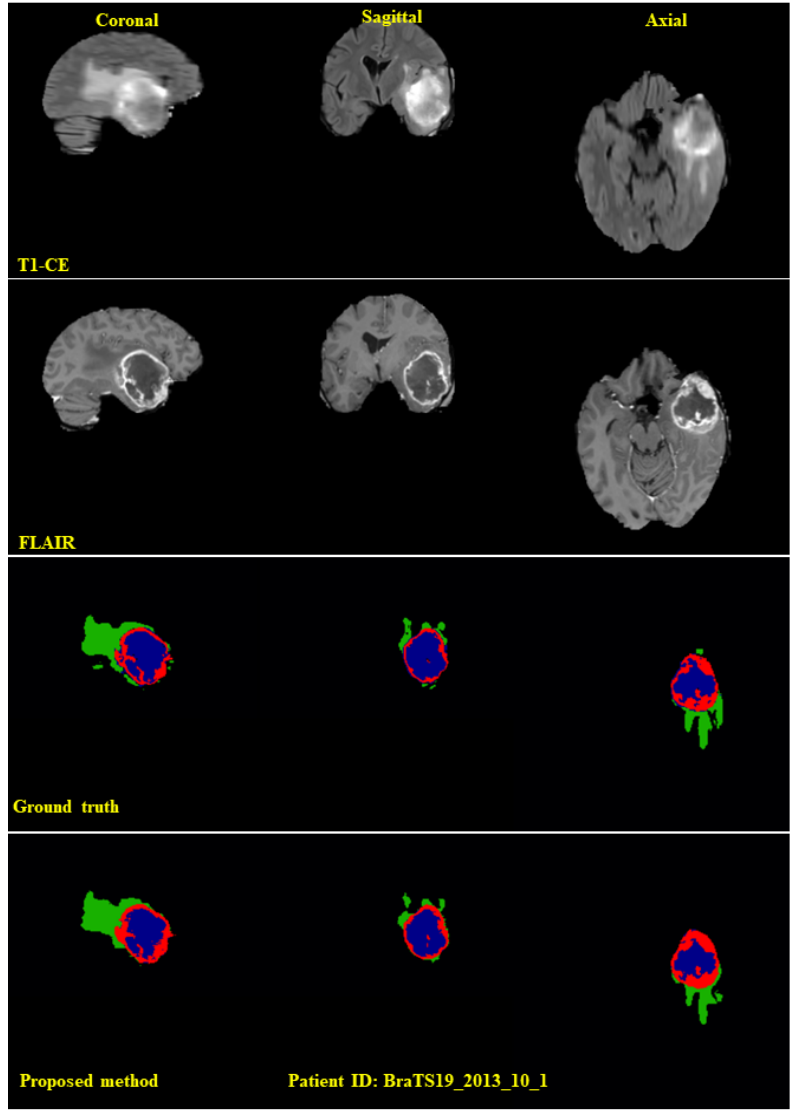

**Fig. 3.** Segmentation results for one sample training data using the proposed multi-resolution model, and comparison with the ground truth. Light blue: necrosis and on-enhancing, green: edema, red: enhancing. The Dice scores reported by the CBICA system for enhancing tumor, tumor core, and the whole tumor are as follows: BraTS19_2013_10_1: 0.82, 0.94, and 0.91.

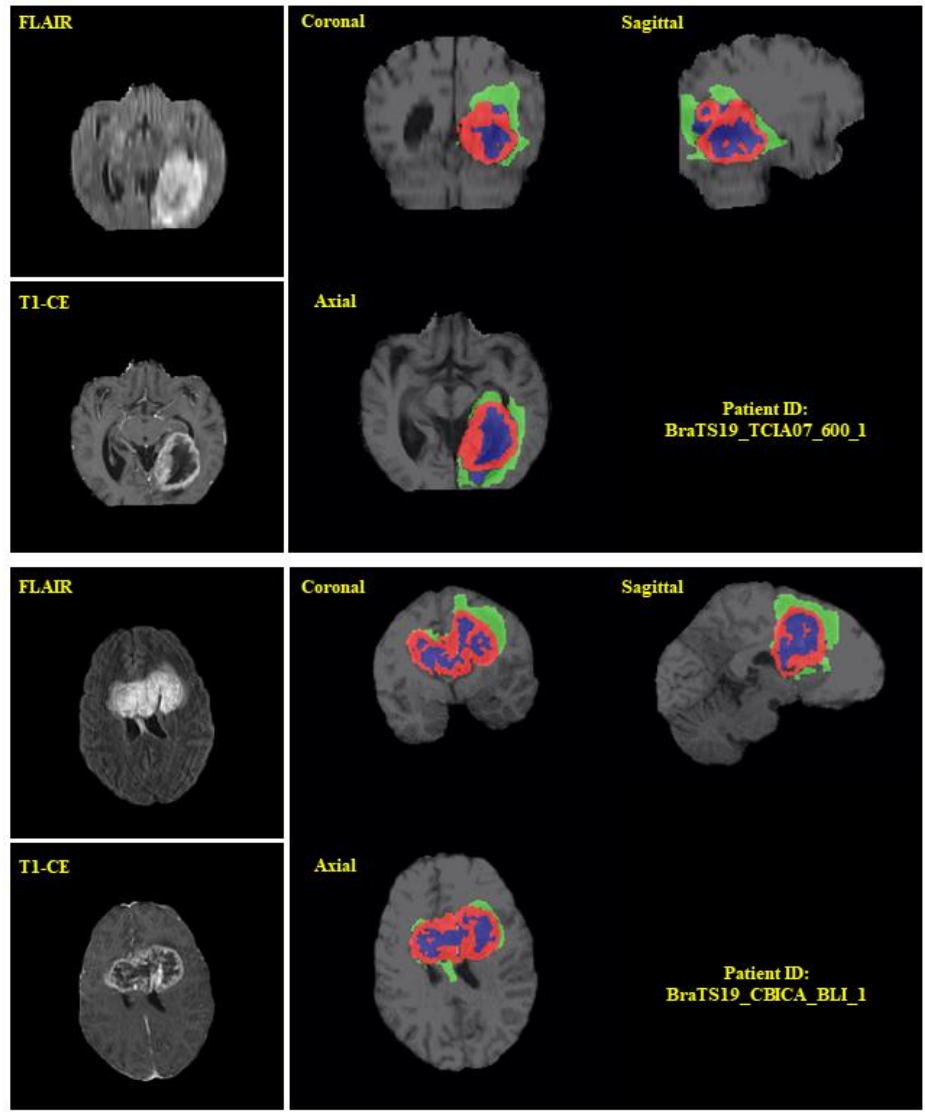

**Fig. 4** Segmentation results for two validation data samples using the proposed multi-resolution model. Light blue: necrosis and on-enhancing, green: edema, red: enhancing. The Dice scores reported by the CBICA system for enhancing tumor, tumor core, and the whole tumor are as follows: Brats19_TCIA07_600_1: 0.86, 0.91, and 0.93; Brats19_CBICA_BLI_1: 0.84, 0.92, and 0.91.

**Table 1.** Segmentation results for validation dataset provided by CBICA portal blind testing system. ET: enhancing tumor, WT: whole tumor, TC: tumor core.

| Dataset | | Dice | | | Sensitivity | | | Specificity | | | Hausdorff (95%) | | |
|---|---|---|---|---|---|---|---|---|---|---|---|---|---|
| | | ET | WT | TC | ET | WT | TC | ET | WT | TC | ET | WT | TC |
| Validation | Mean | 0.71 | 0.84 | 0.74 | 0.68 | 0.82 | 0.74 | 1.00 | 0.99 | 1.00 | 10.11 | 14.00 | 16.06 |
| | STD | 0.26 | 0.13 | 0.22 | 0.26 | 0.14 | 0.22 | 0.00 | 0.01 | 0.01 | 20.39 | 23.44 | 27.64 |
| Test | Mean | 0.70 | 0.82 | 0.72 | - | - | - | - | - | - | 5.59 | 8.42 | 9.14 |
| | STD | 0.23 | 0.18 | 0.29 | - | - | - | - | - | - | 11.64 | 13.22 | 12.68 |

**Table 2.** The results of survival prediction for validation dataset provided by CBICA portal blind testing system.

| Dataset | Accuracy | MSE | Median SE | STD SE | Spearman R |
|---|---|---|---|---|---|
| Validation | 0.52 | 104773 | 37636 | 134872 | 0.20 |
| Test | 0.49 | 408632 | 69696 | 1219534 | 0.28 |

### 3.2 Survival Prediction

In order to validate the survival prediction method, 29 patient cases were specified by the CBICA portal, and the predictions were generated in terms of the number of survival days. Evaluation metrics for this task, were accuracy for classification mode and mean square error (MSE), median and standard deviation of SE, and Spearman R for regression mode. The classes of survival predictions were calculated based on three categories, short (less than 10 months), medium (between 10 to 15 months), and long (more than 15 months). The survival prediction results provided by CBICA system are presented in Table 2.

## 4 Conclusion

In the present study, a three-dimensional multi-resolution learning-based algorithm was proposed in which, instead of patching the image, the whole MR image is passed to the network. The low resolution path was inspired by the U-Net architecture which was modified to take a larger input receptive field and considered the whole input volume rather than partial patches. Although this procedure performed a coarser segmentation, the false positives were successfully eliminated, while the original resolution path produced fine segmentation boundaries. Fusion of these two resolution levels results in increasing accuracy, specificity and sensitivity compared to utilizing each single pipeline separately. The proposed algorithm reached to the Dice scores of 0.84, 0.74 and 0.71 for the whole tumor, core and enhancing tumor on the validation data, and 0.82, 0.72 and 0.70 on the test data set. The hand-crafted statistical and intensity-based features extracted from the segmentation masks are then applied to a random forest classifier for the task of survival prediction. The proposed method acquired MSE and classification accuracy of 104773 and 0.52, for the validation dataset. The corresponding results for the challenge dataset were 408632, and 0.49, respectively.